\documentclass[11pt,emulateapj]{article}
\usepackage[onecolumn]{emulateapj}

\begin{document}

\title{Measuring \lowercase{$\Omega/b$} with weak lensing}
\author{N. Ben\'\i tez \& J. L. Sanz\altaffilmark{1,2}} 
\affil{Astronomy Department, UC Berkeley, 601 Campbell Hall, Berkeley, CA}
\authoremail{benitezn@mars.berkeley.edu}
\altaffiltext{1}{Center for Particle Astrophysics, 
UC Berkeley, 301 Le Conte Hall, Berkeley, CA}
\altaffiltext{2}{On sabbatical leave from IFCA(CSIC-UC), Santander, Spain}

\begin{abstract}

A correlation between the surface density of foreground galaxies 
tracing the Large Scale Structure and the position of distant 
galaxies and quasars is expected from the magnification 
bias effect (Canizares 1981). We show that this foreground--background 
correlation function $w_{fb}$ can be used as a straightforward and 
almost model-free method to measure the cosmological ratio $\Omega/b$. 
For samples with appropriate redshift distributions, 
$w_{fb}\propto \Omega<\delta g>$, where $\delta$ and $g$ are 
respectively the foreground dark matter and galaxy surface density 
fluctuations. Therefore, $\Omega/b \propto w_{fb}/w$, where 
$w\equiv <gg>$ is the foreground galaxy angular two-point 
correlation function, $b$ is the biasing factor, and the proportionality 
factor is independent of the dark matter power spectrum. Simple 
estimations show that the application of this method to the galaxy and 
quasar samples generated by the upcoming Sloan Sky Digital Survey 
will achieve a highly accurate and well resolved measurement of the 
ratio $\Omega/b$.

\end{abstract}

\keywords{Cosmology; gravitational lensing; biasing factor} 

\section{Introduction}

Weak lensing promises to be one of the most effective ways to 
study the properties of Large Scale Structure (LSS) in the next years 
(Kaiser \& Squires 1993; Kaiser 1998; Van Waerbeke, Bernardeau \& Mellier 
1999; Bartelmann \& Schneider 1999). Most of the weak lensing methods 
proposed so far rest heavily on the analysis of background galaxy distorsions 
(Kaiser 1992; Bernardeau et al. 1997, Schneider et al. 1997), 
as they potentially contain finer detail about the LSS than the fluctuations of the 
background number counts, which are affected by intrinsic clustering. However, 
detecting and analyzing the typical shear expected from the LSS offers technical 
problems difficult to underestimate (Kaiser, Squires \& 
Broadhurst 1995; Bonnet \& Mellier 1995; Van Waerbeke et al. 1997; 
Kaiser 1999; Kuijken 1999; although see Schneider et al 1998).
In addition, a number of surveys will be available in the near future, 
as the Sloan digital Sky survey (\cite{sdss}), NOAO (Januzzi et al. 1999) etc. 
with vast amounts of data which may not be optimal for shear analysis due 
to the imaging pixel size or typical seeing. It is therefore necessary to 
develop methods which are able to tap the wealth of cosmological information 
contained in such surveys without relying exclusively on image distorsion analysis. 

Such an approach is provided by the background-foreground correlation 
function $w_{fb}$ (\cite{bar91}; \cite{bar93}). The value of this statistic has 
been calculated by several authors using linear and nonlinear evolution models 
for the power spectrum evolution (Bartelmann 1995; Villumsen 1996; Sanz, 
Mart\'\i nez-Gonz\'alez \& Ben\'\i tez 1997; 
Dolag \& Bartelmann 1997; Moessner \& Jain 1998; Moessner, Jain \& Villumsen 1998). 
The two most obvious cases in which it is possible to measure $w_{fb}$ are 
galaxy--galaxy correlations and quasar--galaxy correlations. The detection of the 
latter has a long and controversial history (for a discussion see Schneider, 
Ehlers \& Falco 1992, Ben\'\i tez 1997), but it seems to be already well 
established. However, due to the scarcity of complete, well defined quasar 
catalogs not affected by observational biases, the results have low 
signal--to--noise and are difficult to interpret 
(Ben\'\i tez, Sanz \& Mart\'\i nez--Gonz\'alez 1999).  
The value of the expected amplitude for the low-z galaxy--high-z galaxy 
cross--correlation is rather small, and hard to measure within typical single CCD 
fields (\cite{vil97}). Only with the advent of deep, 
multicolor galaxy samples and reliable photometric redshift techniques 
it has been possible to detect this effect (Herranz et al. 1999). 

To interpret the measurements of $w_{fb}$ using the calculations mentioned above, 
it is necessary to assume a certain shape for the power spectrum. 
Unfortunately it is still far from clear whether the most popular ansatz, 
that of Peacock and Dodds (1996)---or any other for that matter---provides 
an accurate fit to the LSS distribution (see e.g. Jenkins et al. 1998). 
It is thus desirable to develop methods whose application is not 
hindered by this uncertainty. An example is the statistic $R$ 
(Van Waerbeke, 1998), which combines 
shear and number counts information, and can be used to measure the scale dependence 
of the bias. The value of $R$ is almost independent of the shape of the 
power spectrum if the foreground galaxy distribution has a narrow 
redshift range. Here we show that something similar can be achieved with 
$w_{fb}$, with the additional advantage of being able to do without 
the shear information in those cases where the latter is difficult to obtain.
 
It follows from the magnification bias effect (Canizares 1981, Narayan 1989, 
Broadhurst, Taylor \& Peacock 1995) that the surface number density of 
background population $n_b$ 
is changed by the magnification $\mu$, associated with a foreground 
galaxy population with number density $n$ as 
$g_b =(\alpha-1)\delta \mu$, where $g_b$ is the 
perturbation in the galaxy surface density $n_b$ ($g_b=n_b/<n_b>-1$), 
$\alpha$ is the logarithmic slope of the number counts and in the weak 
lensing regime $\mu \approx 1+ \delta\mu$ , $\delta\mu << 1$. 
Since $\mu\approx 1+2\kappa$ and $\kappa$, the convergence, is proportional 
to the projected matter surface density $\Sigma$, it follows that 
$g_b \propto \delta\kappa \propto \delta\Sigma \propto \Omega\delta$, 
where $\delta$ is the dark matter surface density perturbation. 
Therefore $w_{fb} \equiv <gg_b>\propto \Omega <g\delta>$
and assuming linear and deterministic bias $w_{fb}\propto\Omega b^{-1}w$
where $w$ is the two-point galaxy correlation function for the background 
populations, and the biasing factor $b$ is defined by $w=b^2w_{\delta\delta}$. 
This result provides a straightforward, virtually 
model independent method to estimate the ratio $\Omega/b$. 
Williams \& Irwin 1998 arrived to a similar expression using a phenomenological 
approach. The redshift distorsion method 
provides the quantity $\beta\propto \Omega^{0.6}/b$ (Dekel 1994), so combining 
both one can estimate $\Omega$ and $b$ separately. 

The outline of the paper is the following. In Sec. 2 we show rigorously 
that under reasonable assumptions $w_{fb}\propto \Omega b^{-1}w$. 
Sec. 3 explores the application of this method to future and ongoing surveys and 
Sec. 4 summarizes our main results and conclusions. 

\section{Foreground-Background correlations}

Let us consider two populations of sources: a background one (e.g. quasars or 
galaxies) and a foreground one (e.g. galaxies), placed at different distances 
$\lambda$ with p.d.f.'s $R_b(\lambda)$ and $R(\lambda)$, respectively. In 
Sanz et al. (1997), the background-foreground correlation $w_{fb}$ was 
calculated as a functional of the power spectrum $P(\lambda , k)$ of 
the matter fluctuations at any time ($\lambda$ is the comoving distance 
from the observer to an object at redshift $z$, 
$\lambda=\frac{(1 + \Omega z)^{1/2}-1}{(1 + \Omega z)^{1/2}-1+\Omega}$ and 
$\Omega<1, \Lambda=0$ ): 

\begin{equation} 
w_{fb}(\theta ) = ({\alpha}_b - 1)C_{\mu \delta}(\theta ), 
\end{equation} 

\begin{equation} 
C_{\mu \delta}(\theta ) = 12\Omega\int_0^1d\lambda \,b(\lambda, s\theta )T_b(\lambda ) 
R(\lambda ){\lambda^2\over(1 - \lambda)^2}\tau(\lambda, s\theta ), 
\end{equation} 

\begin{equation} 
\tau(\lambda, s\theta ) \equiv \frac{1}{2\pi}\int_0^{\infty}dk\,kP(\lambda , k) 
J_0(ks\theta ) ,\ \ \ s\equiv \frac{\lambda}{1 - (1 - \Omega ){\lambda}^2}, 
\end{equation} 

\noindent where $b(\lambda,s\theta )$ is the bias factor for the foreground galaxies  
(assumed to be linear, non-stochastic but possibly redshift and scale dependent), 
${\alpha}_b$ is the slope of the background source number counts and 
$C_{\mu \delta}(\theta)\equiv 2<\mu (\vec{\phi}) 
\delta(\vec{\phi} + \vec{\theta})>$ is the correlation between the magnification and 
the mass density fluctuation. 
$T_b(\lambda)$ is the lensing window function 
given by equation (7) in Sanz et al. (1997): \hfill\break 
$T_b(\lambda )\equiv \frac{1}{\lambda }\int_{\lambda}^1\frac{du}{u}R_b(u) 
(u - \lambda )\frac{1 - (1 - \Omega )u\lambda }{1 - (1 - \Omega ){\lambda }^2}$. 

The angular two point correlation function for the foreground population $w$ 
can be obtained as 

\begin{equation}
w(\theta) = \int_0^1d\lambda \,b^2(\lambda,s\theta )R^2(\lambda )
[1 - (1 - \Omega ){\lambda}^2]\tau(\lambda, s\theta ),
\end{equation}

\noindent where $w(\theta )\equiv <g(\vec{\phi})g(\vec{\phi} + \vec{\theta})>$ 
is the foreground angular galaxy--galaxy correlation function.

If we assume that the foreground and background galaxies are 
concentrated at the `effective' distances $\lambda_f$ and $\lambda_b$
(a good approximation for realistic, nonoverlapping redshift distributions, 
Sanz et al. 1997), the previous formulas can be rewritten as
\begin{equation}
w_{fb}({\lambda}_f,{\lambda}_b, \theta ) \simeq (\alpha_b-1)12\Omega \frac{2b}
{a_f^2{\Sigma}_c}\tau ({\lambda}_f, s_f\theta )
\end{equation}

\begin{equation}
w({\lambda}_f, {\Delta \lambda}_f, \theta) \simeq b^2\frac{1}
{\Delta \lambda_f}[1 - (1 - \Omega){\lambda_f}^2]
\tau ({\lambda}_f, s_f\theta ),
\end{equation}

\noindent where ${\Sigma}_c({\lambda}_b, {\lambda_f}) \equiv 
\frac{2D_b}{D_fD_{fb}}$ is the critical surface mass distance defined by the 
angular distance $D$, $a_f\equiv \frac{{(1 - {\lambda}_f)}^2}{1 - (1 - \Omega){\lambda}^2_f}$ is the scale factor, and 
${\Delta \lambda}_f\simeq 1/R_f({\lambda}_f)$ is
the width of the p.d.f. 
$R_f(\lambda)$ (e.g. the FWHM for a Gaussian distribution). Dividing Eq. (5) by Eq. (6) and simplifying one obtains 

\begin{equation}
\frac{w_{fb}} 
{w}
\simeq Q\Delta z_f(\alpha_b - 1){\Omega \over b} 
\label{eq1}
\end{equation}
\noindent i.e. the two correlations are proportional. The 
proportionality factor $Q$ has the form 
\begin{equation}
Q\equiv 
\frac{6\lambda_f(1 - \lambda_f/\lambda_b)[1 - (1 - \Omega )\lambda_f\lambda_b]
(1-\lambda_f)}
{[1 - (1 - \Omega )\lambda_f][1 - (1 - \Omega )\lambda^2_f]^2}\label{q}
\end{equation}

Fig. 1 (top) shows the dependence of $Q$ on $z_b$ and $z_f$ for the $\Omega=0.3, 
\Lambda=0$ case. For the $\Omega+\Lambda=1, \Lambda>0$ case, there is not an explicit form for 
the angular distances, and the corresponding expression for $Q$ is (Fig. 1, bottom)
\begin{equation}
Q=\frac{12(1+z_f)}{\Sigma_c(z_f,z_b)\sqrt{1+\Omega z_f+\Lambda[(1+z_f)^{-2}-1]}}
\end{equation} 

Note in Fig. 1 that for $z_f<0.2$ and $z_b>1$ the value of $Q$ in both open and 
flat geometries changes very slowly with $z_b$ and almost linearly 
with $z_f$. To better understand this behavior from Eq. (\ref{q}) let's assume that 
$z_f<<1<z_b$. In that case, ${\lambda}_f \simeq s_f \simeq \frac{1}{2}z_f$, 
$a_f\simeq 1 - z_f$ and $Q\approx 3 z_f $. Therefore, $w_{fb}$ 
is almost independent of $\lambda_b$ (Fig. 1) and roughly 
$w_{fb}\simeq 3z_f{\Delta z}_f(\alpha_b-1)\Omega b^{-1}w$~. 
Fig. 2 also shows that for the same redshift range, the amplitude of $Q$ is also 
approximately independent of $\Omega$, allowing an empirical, model--independent 
estimation of $\Omega/b$. 

\section{Practical application}

The main quantity which determines the signal in the measurement of the 
angular correlation function is the excess (or defect) in 
the expected number of galaxy pairs. For a bin with surface $A_{bin}$ at 
distance $\theta$ this number will be 
\begin{equation}
\Delta N(\theta) \approx (\pm)N_f n_b A_{bin}(\theta)w_{fb}(\theta)
\end{equation}
Therefore, the signal--to--noise of the detection will be roughly 
\begin{equation}
{S \over N}\sim {\Delta N \over \sqrt{N}}=\sqrt{N_f} \sqrt{n_b A_{bin}} w_{fb}(\theta)
\end{equation}
We have not included the scatter due to the clustering of foreground galaxies 
and background galaxies, but it is unlikely that this effect will increase the 
noise over Poisson more than a factor of a few for any realistic case. 
The quantity $n_b A_{bin}$ is the number of background galaxies 
per bin. If we assume radial concentric bins of width $\Delta \theta$ and 
$w_{fb}=A_{fb}\theta^{-\gamma}$, the signal to noise in each bin will be
\begin{equation}
{S \over N}\sim A_{fb}\sqrt{2\pi N_f n_b\Delta\theta} \theta^{0.5-\gamma}
\end{equation}
Since $\gamma$ is typically $0.7-0.8$, the efficiency of the method will decrease 
very slowly with radius, allowing to map $\Omega/b$ up to very large scales.

The SDSS will obtain $\sim 10^6$ spectra for a population of $<z>\sim 0.1$ galaxies 
(\cite{sdss}), forming a splendid foreground sample. 
It may be assumed that its projected angular correlation function will 
be similar to that of the APM catalog, which has an amplitude 
$A=0.44$ at $1'$ and a slope $\gamma=0.668$ (\cite{apm}). 
The SDSS will also obtain $S/N\sim10$ $u'g'r'i'z'$ 
photometry for $10^8$ galaxies in a $10^4$ degree region, for which 
photometric redshifts will be estimated. The background sample can be 
formed by those galaxies with $z>0.4$ and $<z>\approx 0.7-0.8$, with 
$n_b\approx 1-1.5/\sq\arcmin$. The value of $Q$ cannot be considered 
constant for these 
two samples (see Fig. 1) because of the range of redshifts involved. Luckily, the existence 
of spectroscopic redshifts for the foreground sample will allow to select 
subsamples with an extremely thin redshift distribution for which $Q$ is approximately 
constant, each yielding independent estimates of $\Omega/b$ which can afterwards be combined 
together. A rough estimate of the expected result using Eq. (12) gives 
\begin{equation}
\left({S\over N}\right)_{SDSS}\sim 20{\Omega \over b}\sqrt{\left({\Delta\theta \over 5'}\right)} 
{\theta}^{-0.2} 
\end{equation}
(we have assumed $N_f= 10^6, n_b=1.5, \Delta z_f=0.05 , Q=0.25, \alpha_b-1=0.5$)
This shows that it will be possible to map $\Omega/b$ with an extremely good combination 
of resolution and accuracy using the SDSS. At large radii, the bias factor is expected 
to become constant (\cite{coles}), which means that the bin size can be made as large 
as desired without being affected by the scale dependence of $b$ and therefore 
substantially decrease the error in the estimation of $\Omega/b$. 
It is obvious from the above numbers that the main source of errors 
will not be the shot noise, but contamination due to low redshift galaxies which may sneak 
into the high--redshift sample 
creating a spurious correlation. This contamination can be very effectively minimized by 
applying a Bayesian threshold (Ben\'\i tez 1999). 
In addition, it will be possible to accurately quantify and correct this effect using a 
calibration sample with spectroscopic redshifts. 

  The SDSS also plans to obtain spectra for a sample of $10^5$ red luminous 
galaxies with $<z>\sim 0.4$ and $10^5$ QSOs. Due to the much lower density 
of the background sample, it will not be possible to attain such a good spatial 
resolution in the measurement of $\Omega/b$, but these samples will allow to trace 
the evolution of biasing with redshift up to $z\sim 1$. However, experience shows 
that in this case the utmost attention will have to be paid to eliminating or 
accounting for the observational biases in the QSO detection and identification 
procedure, which can totally distort the estimation of $w_{bf}$ (Ferreras, 
Ben\'\i tez \& Mart\'\i nez-Gonz\'alez 1997, Ben\'\i tez, Sanz \& 
Mart\'\i nez-Gonz\'alez 1999). 

\section{Conclusions}

A correlation (positive or negative) between the surface density of 
foreground galaxies tracing the Large Scale Structure and the position 
of background galaxies and quasars is expected from the magnification 
bias effect (Canizares 1981). We show that this foreground--background 
correlation function $w_{fb}$ can be used as a straightforward 
and almost model-free method to measure the cosmological ratio $\Omega/b$. 

For samples with appropriate redshift distributions, 
$w_{fb}\propto \Omega <\delta g>$, where $\delta$ and $g$ are 
respectively the foreground dark matter and galaxy surface density 
fluctuations. Therefore, $\Omega/b \propto w_{fb}/w$, 
where $w\equiv <gg>$ is the foreground galaxy angular two-point 
correlation function, $b$ is the biasing factor, and the proportionality 
factor is independent of the dark matter power spectrum. 

Simple estimations show that the application of this method to the galaxy and 
quasar samples generated by the upcoming Sloan Sky Digital Survey 
will achieve a highly accurate and well resolved measurement of the 
ratio $\Omega/b$.

\acknowledgments

The authors thank Enrique Mart\'\i nez--Gonz\'alez for interesting discussions. 
NB acknowledges a Basque Government postdoctoral fellowship and partial 
financial support from the NASA grant LTSA NAG-3280. JLS acknowledges a 
MEC fellowship and partial financial support from CfPA.

\begin{figure}[t] 
\epsscale{.65}
\plotone{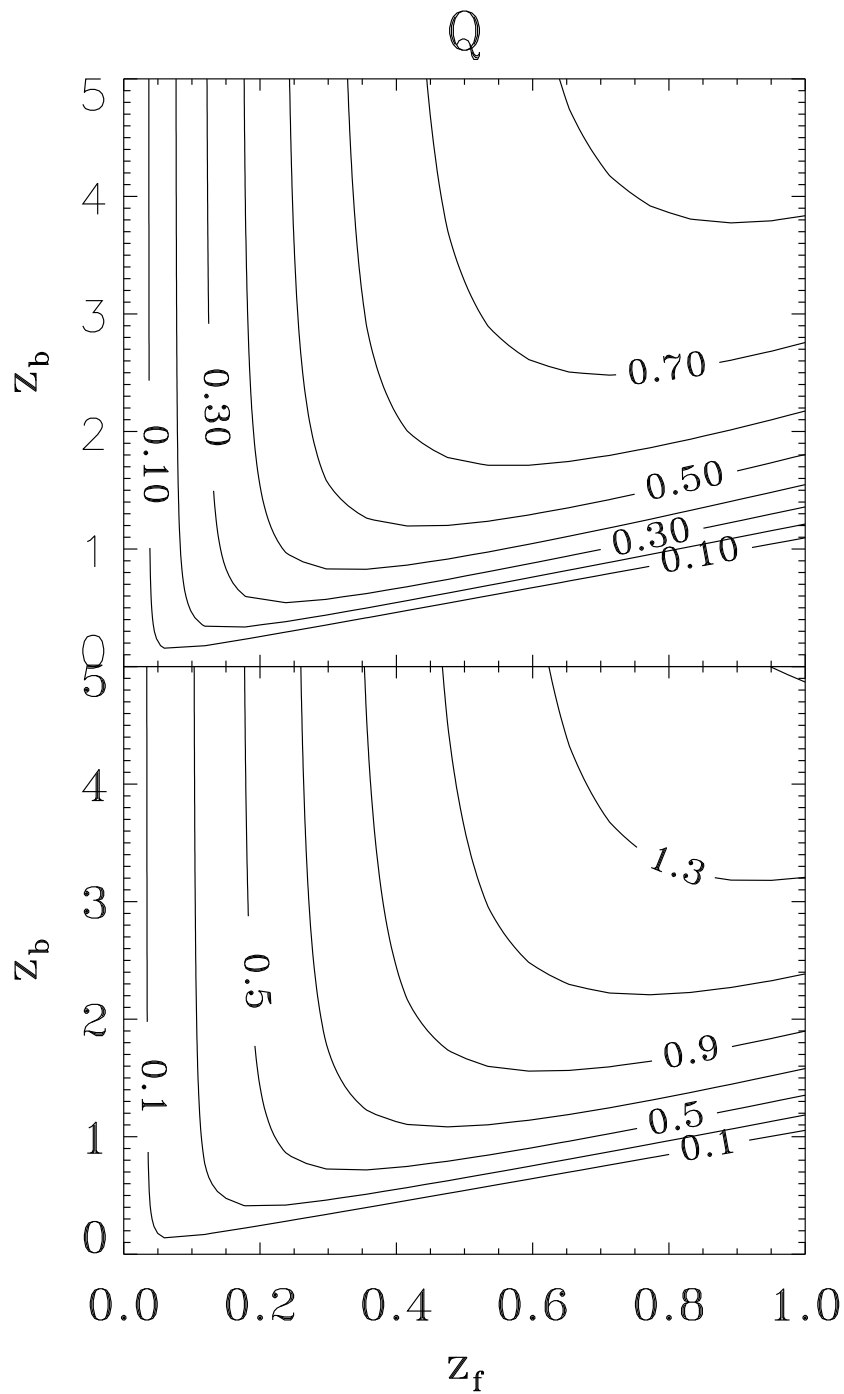}
\caption{Dependence of the factor $Q$ (Eqs. \ref{q}, 11) on the redshift 
of the foreground and background samples for open 
(top, $\Lambda=0$) and flat (bottom, $\Omega+\Lambda=1$) geometries 
($\Omega=0.3$ in both cases).}
\end{figure}

\begin{figure}[t] 
\epsscale{.65}
\plotone{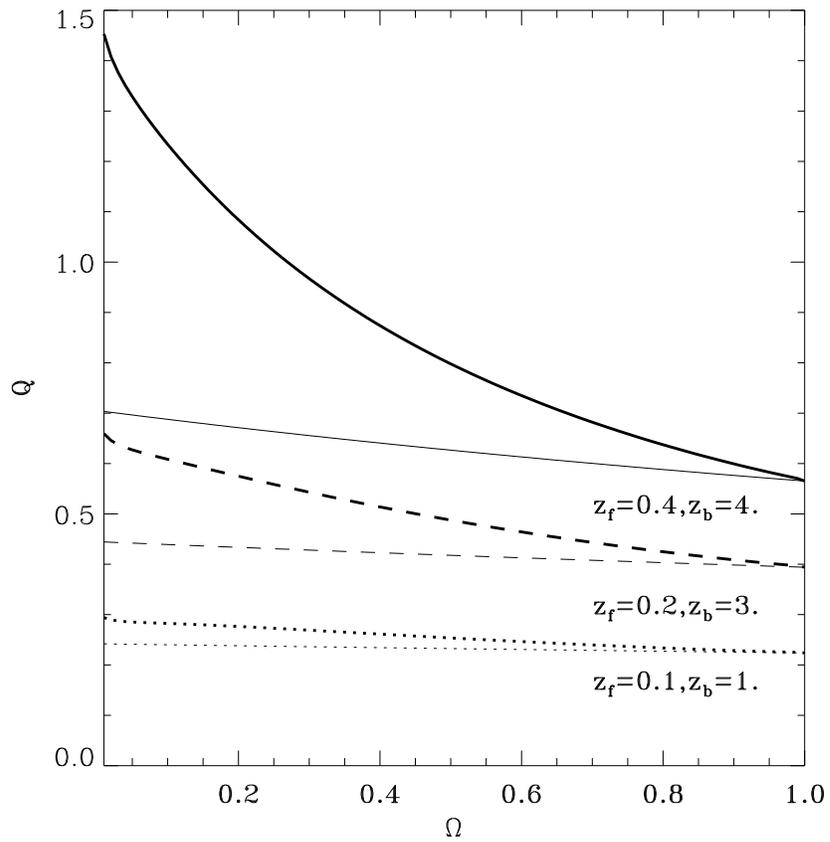}
\caption{Dependence of the factor $Q$ (Eqs. \ref{q}, 11) on $\Omega$ for 
different redshifts of the foreground and background populations 
$z_f$ and $z_b$. The thick lines correspond to $\Omega+\Lambda=1$ universes 
and the thin ones to $\Lambda=0$.}
\end{figure}

\end{document}